\documentstyle[aps,psfig,twocolumn]{revtex}
\begin{document}
\draft
\flushbottom
\twocolumn[
\hsize\textwidth\columnwidth\hsize\csname @twocolumnfalse\endcsname
 
%\baselineskip=24pt 
%\bibliographystyle{unsrt}
%\newcounter{chapter} 
%\newcounter{figno}[chapter] 
%\setcounter{figno}{0} 

\title{Collapse of the charge ordering gap of 
Nd$_{0.5}$Sr$_{0.5}$MnO$_{3}$ in an applied magnetic field} 
\author{Amlan Biswas$^1$ ~\cite{ADDRESS}, 
Anthony Arulraj$^2$, A. K. Raychaudhuri$^1$ ~\cite{NPL} and C. N. R. Rao$^2$}
\address{$^1${\it Department of Physics, 
Indian Institute of Science, Bangalore-560012,
INDIA}\\
$^2${\it Solid State and Structural Chemistry Unit, Indian Institute of Science,
Bangalore-560012, INDIA}}
\date{\today}
\maketitle
\tightenlines
\widetext
\advance\leftskip by 57pt
\advance\rightskip by 57pt

\begin{abstract}

 We report results of tunneling studies on the charge ordering compound
  Nd$_{0.5}$Sr$_{0.5}$MnO$_{3}$ in a magnetic field up to 6T and for       
 temperature down to 25K.We show that a gap (2$\Delta_{CO}$$\approx$0.5eV
  opens up in the density of 
state (DOS) at the Fermilevel ($E_F$) on charge ordering ($T_{CO}$=150K) 
which 
collapses 
in an applied magnetic field when the charge ordered 
state melts.  There is
a clear correspondence between the behavior of the resistivity and the gap 
formation and its collapse
in an applied magnetic field.
 We conclude that a 
gap in the DOS at $E_F$ is necessary for the stability of the charge
ordered state.

\end{abstract}
]
\narrowtext
\tightenlines

% \begin{multicols}{2}
%\newpage

The phenomenon of Colossal Magnetoresistance (CMR)
in hole-doped rare-earth manganites has, in recent times, been the subject 
of intense research efforts ~\cite{RVH,CHAH,URUSHI,MAHI1}. 
These oxides belong to the ABO$_3$ class of perovskite oxides and 
contain a mixed  
valency of Mn, which occupies the B site. The A site is occupied by 
rare-earth 
ions like Nd, La, Pr etc. or by divalent ions like Pb, Ca, Sr etc. These 
oxides, 
therefore, 
have the general formula $R_{1-x}M_x$MnO$_3$, where $R$ is 
La, Nd etc. and $M$ is Ca, Sr, Pb etc. One of the most interesting
properties of these materials is the insulator to metal transition which 
occurs when these materials are cooled below their ferromagnetic Curie 
temperature ($T_c$).
The stability of this
ferromagnetic metallic state depends on the size of the A-site cation.
When 
smaller
ions are substituted in the A-site, it lowers the
ferromagnetic $T_c$ and makes the ferromagnetic 
state unstable at low $T$~\cite{BELL,MAHIY}.
For certain values of $x$ (particularly when $x \approx$ 0.5) the
unstable ferromagnetic state can make a transition to an
insulating state when cooled much below $T_c$ ~\cite{TOKPH}. This phenomenon 
is called charge ordering (CO). 
This occurs due to the real space ordering of
Mn$^{3+}$ and Mn$^{4+}$ ions in alternate sublattices. This
transition is also associated with large lattice
distortions~\cite{TOKPH,BRIT,NDCA}. For the
particular solid studied, i.e. 
Nd$_{0.5}$Sr$_{0.5}$MnO$_{3}$, the charge
ordering at temperature $T_{CO}$ is accompanied by a spin ordering
to an antiferromagnetic state. The type of charge ordering and
spin ordering observed experimentally 
depends on the radius of the $R$ and $M$
cations.

It has been shown in an earlier report that a charge ordering gap opens up
in the density of states (DOS) near the Fermi level ($E_F$) of these 
materials, when the sample is cooled below $T_{CO}$ ~\cite{amlanco}. The   
formation of the gap was detected through tunneling studies using a Scanning 
Tunneling Microscope (STM). We
call this  gap the CO
gap, $\Delta_{CO}$. In Nd$_{0.5}$Sr$_{0.5}$MnO$_{3}$ 
the limiting  value of $\Delta_{CO}(T \rightarrow 0)$
was measured to be about 0.25eV and  
$\Delta_{CO}\rightarrow$0 as $T\rightarrow T_{CO}$ ~\cite{amlanco}. 
Photo electron   
spectroscopy also detected a gap in the DOS at $E_F$ of similar 
magnitude for the compound Pr$_{0.5}$Sr$_{0.5}$MnO$_3$ 
~\cite{ashish}. The value of the gap
is rather large compared to the temperature at which the CO takes place. In 
fact 2$\Delta_{CO}$/T$_{CO}$$\approx$40. Interestingly, the value of 
$\Delta_{CO}$ is comparable to the scale of nearest neighbor Coulomb 
interaction and a recent theory predicts formation of such a gap on CO 
transition ~\cite{MIS}.

The charge-ordered 
state in materials like Nd$_{0.5}$Sr$_{0.5}$MnO$_{3}$ is unstable in a 
magnetic field. The applied field can  ``melt'' the CO state leading to the
re-emergence of the ferromagnetic metallic state. 
The field needed to melt the 
CO state depends on the temperature. Details can be 
found in reference ~\cite{TOKPH,kumar}.

In this paper we address an important question: is it necessary to
have a gap in the DOS at $E_F$ to ensure stability of the CO state?
We introduce instability in the CO state by application of a magnetic field
and follow the behavior of $\Delta_{CO}$.
The results presented here are the
\begin{figure}
\centerline{
\psfig{figure=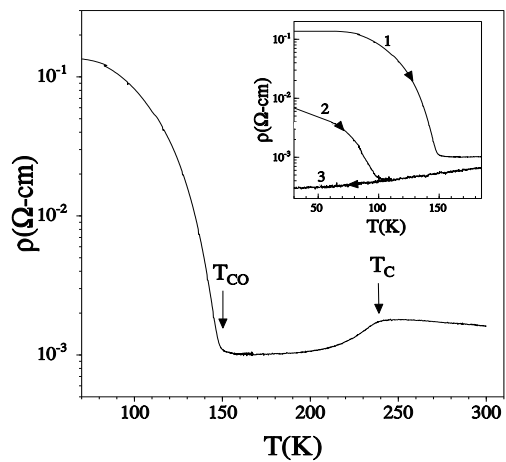,width=7.0cm,height=6.5cm,clip=}
}
\caption{The $\rho$ vs. $T$ plot for Nd$_{0.5}$Sr$_{0.5}$MnO$_{3}$
for a zero field warming cycle on a zero field cooled
sample. The inset shows
the effect of a magnetic field on the resistivity behavior of
Nd$_{0.5}$Sr$_{0.5}$MnO$_{3}$. The arrows indicate the direction of the
change in temperature. The zero
field cooled sample was first warmed in zero field
(curve 1), then the zero
field cooled sample was warmed in a 6 T field (curve 2) and then the
sample was field cooled in 6 T (curve 3).}
\end{figure}

\begin{figure}
\centerline{
\psfig{figure=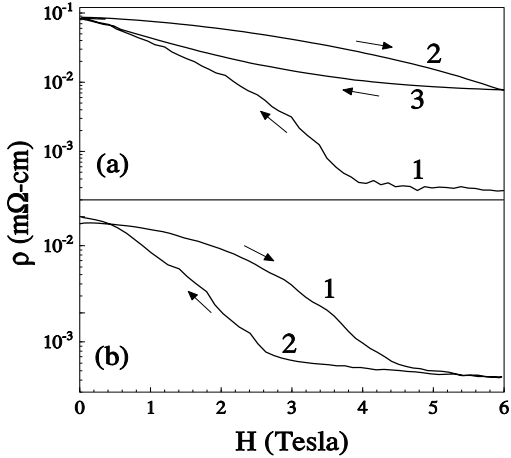,width=7.0cm,height=6.5cm,clip=}
}
\caption{The $\rho$ vs. $H$ plots at two temperatures (a) 4.2 K and
(b) 112 K. In (a), the sample was field cooled
to 4.2 K in 6 T. The sequence of the
the $\rho$ vs. $H$ data taken thereafter is marked by the numbers. The
sample was then heated in zero field to 112 K and another set of $\rho$ vs.
$H$ data were taken. The sequence is marked by the numbers on the curves.}
\end{figure}
first
tunneling studies carried out in the CO state in the presence of a magnetic
field. We have carried out the
experiment using a variable temperature STM
on the
widely studied CO compound
Nd$_{0.5}$Sr$_{0.5}$MnO$_{3}$. We make
the following three main
observations:
(a) Application of
the magnetic field reduces $\Delta_{CO}$ and
for a high enough magnetic field this CO gap collapses
($\Delta_{CO}$$\rightarrow$0),
(b) The magnetic field at which the CO gap collapses depends on the
$T/T_{CO}$ ratio and closely follows the
$H-T$ phase diagram obtained from the resistivity data.
(c) For $T/T_{CO} << 1$ there is a distinct signature of a
two phase state, when the magnetic field is varied and the CO state melts.

The sample used in this work has been prepared by the solid state 
method and is polycrystalline in nature. The sample was characterized by  
x-ray and titration.  
A detailed description 
of the sample preparation and characterization 
is given in reference ~\cite{MAHI1}. The resistivity ($\rho$)
as a function of 
temperature is shown in figure 1.
The $\rho$ vs. $H$ curves for two temperatures are shown in figure 2.
We find that the $T_c$ = 240K 
and $T_{CO}$ = 150K which 
are close to the values $T_{c}$ =255K and $T_{CO}$ = 158K
found for the single crystalline samples of the same material
~\cite{TOKPH}.
The $\rho$ of the single crystal for $T > T_{CO}$ is $\sim$ 10$^{-3}$ 
$\Omega$ cm which is similar to our sample. However,
the jump in $\rho$ for
the single crystal when $T$ goes below $T_{CO}$, is about 4 orders of 
magnitude ~\cite{TOKPH}. Also, the transition in $\rho$ at 
$T \approx T_{CO}$ is much sharper for the single crystal.
The somewhat smeared
nature of the transition for the polycrystalline sample (mostly arising
due to disorder) will not
affect the determination of the gap 
$\Delta_{CO}$ for $T \ll T_{CO}$ but it can
affect the value of $\Delta_{CO}$ close to $T_{CO}$.

The tunneling spectroscopy (TS) investigation 
was carried out with a home made variable temperature 
high vacuum STM, using a 
platinum-rhodium tip~\cite{amlanco,amlanprb}.
The tip-sample separation
(tunneling gap), for a fixed bias, was kept constant
using a feed back loop. For a given set of the tip and 
the sample, the tip-sample
distance determines the value of the tunneling current ($I_t$) for
a given bias. 
In this experiment, at each temperature 
$I_t \approx$ 1.5 nA was first established at a 
bias $\approx$ 1.4 V. 
Since the observed $\Delta_{CO}$ is less than 1 eV, this ensured
that the tip did not crash against the sample in order to maintain a
constant current when a gap opens
up below $T_{CO}$. The details of the experimental procedure for $I-V$ data 
acquisition are described elsewhere 
~\cite{amlanco,amlanprb}. The tunneling spectra for $T=109$ K and for
different field values are shown in figure 3. 
The gap which opens up at $E_F$ 
(which corresponds to the zero bias) for $T < T_{CO} = 150$K
can be clearly seen for the spectra taken at low magnetic fields
and is marked in the figure as 2$\Delta_{CO}$.

The $\rho$ vs. $T$ curve in a field of 6 T is shown in 
the inset of figure 1 along with the zero field data. 
The observed results depend on whether 
the data have been taken for field cooled (FC) or zero field cooled (ZFC) 
samples. The sequence of the change in $T$ and $H$ is mentioned in the 
figure caption. The applied magnetic field on the ZFC sample shifts the
$T_{CO}$ to lower temperatures (curve 2).
Curve 3 shows the data 
on the FC sample where the CO does not set in down to 4K in a field of 6T. 
In figure 2 we show the $\rho$ vs $H$ data at two temperatures $T$=4.2K 
($T/T_{CO}$ =.028) and 112K ($T/T_{CO}$ = 0.747). The data are taken for an 
FC sample.The $H-T$ phase diagram obtained by us agrees 
qualitatively with the $H-T$ diagram 
obtained for the single crystal of the same material
~\cite{TOKPH}. The measurement of
 $\Delta_{CO}$ 
in a magnetic field follows the same history as in the resistivity measurement. 

Steps followed for measuring the variation of $\Delta_{CO}$ with the 
applied magnetic field are described below. The temperatures at which the 
variation of 
$\Delta_{CO}$ with $H$ are measured, are chosen carefully so that different 
regions of the $H-T$ plane are probed. We mainly probed 
two regions in temperature, 
one deep inside the CO state ($T \ll T_{CO}$) and other close to $T_{CO}$.
The temperature is then stabilized 
at one such chosen temperature. The tunneling current is established as 
described earlier and the $I-V$ curves are recorded at different values of 
$H$. From the $I-V$ curves the $G-V$ curves are obtained. 
To move in the other direction of the $H-T$ phase diagram, i.e. to 
measure the change of $\Delta_{CO}$ with temperature in the presence of a 
constant magnetic field, the field was set at a particular value and the 
temperature was stabilized at different values where the $I-V$ data were 
acquired as described earlier. For all these measurements
the direction and
sequence of the change in $T$ and $H$ was chosen carefully to account for
the history dependence of the charge ordering transition.

In figure 3 shows a set of $I-V$ and $G-V$ curves taken on a
ZFC sample.
The sample was cooled to 109K ($T/T_{CO}$=0.71) in
zero field.
After the temperature stabilization the magnetic field was ramped up to 4T
and
down to zero again.  One can clearly see from the tunneling curves that the
gap in DOS at $E_F$ in zero field collapses 
in a field of 4T and reappears again when the
field is reduced. This variation of $\Delta_{CO}$ with $H$
is shown in figure 4.
For field ramped up and down, we observe nearly the same
\begin{figure}
\centerline{
\psfig{figure=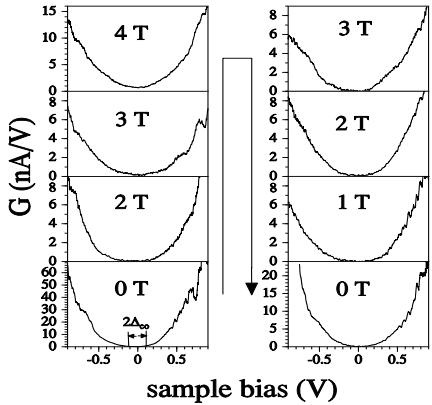,width=8.0cm,height=6.8cm,clip=}
}
\caption{
The evolution of $G-V$ curves with magnetic field at 109 K. 2$\Delta_{CO}$
is marked for one curve.
The sequence of the experiment is marked by the arrow.}
\end{figure} 
\begin{figure}
\centerline{
\psfig{figure=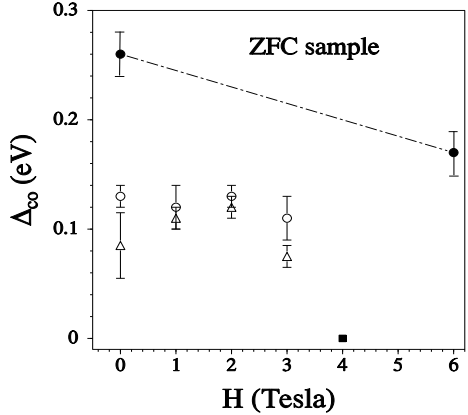,width=7.0cm,height=6.5cm,clip=}
}
\caption{
The variation of $\Delta_{CO}$ with field at 109 K. The open
triangles denote the field increasing run and open circles are the data
points obtained during the field decreasing run. The filled square
denotes the collapse of $\Delta_{CO}$ at 4 T. The filled circles are the
values of $\Delta_{CO}$ for T=34 K.}
\end{figure} 
value for $\Delta_{CO}$. The
corresponding $\rho$ vs. $H$ curve is shown in figure 2 where one 
sees a ``hysteresis'' when field is ramped up and down. In the field 
increasing cycle the $\rho$ collapses at 4.5T while in field reducing cycle
the CO state reappears at 2.5T.In case of $\Delta_{CO}$ the collapse and the 
onset of the gap takes place at the same field of $H \approx$ 4T. 
In figure 4 we 
also show the variation of the gap for a ZFC sample at 34K. Even at a field 
of 6 T, though the gap is reduced substantially, it does not collapse. This 
corresponds to the resistivity data on a ZFC sample. Probably at 
a higher field than accessible to us we can melt the CO gap at low    
temperatures.

The data on FC samples are shown next. In figure 1 it can be seen that for  
the sample field cooled in 6T there is no CO state. We also find that no  
gap opens up in the DOS at $E_F$ in a field of 6T down to 4.2K in 
an FC sample. But we 
make a very interesting observation regarding the zero bias conductance   
$G_0$ which shows a gradual a decrease in a 6T field as $T$ is reduced     
below 150 K (the zero field $T_{CO}$).
The 
zero 
bias 
conductance,$G_0$ is a  
proportional measure of the DOS at $E_F$, $N(E_F$). In order to compare 
it at different $T$ we normalize it by the tunneling conductance for a bias     
value far 
away from the Fermi level (in our case $V$=0.9V).
 The normalized tunneling 
conductance $G_0/G_{0.9}$ is shown as a function of $T$ in a field of 6T 
for the field cooling run, in figure 5. We find that though the CO gap
does not appear even down to 4.2K, there 
is a big reduction in the normalized DOS at $E_F$ on cooling. The DOS 
is small but finite at the lowest T. The temperature dependence 
of $G_0/G_{0.9}$ 
down
to $T=120$K is such that a smooth extrapolation
(shown by the dotted line in figure 5) will
give zero DOS at $E_F$ at T=100K which is
the CO temperature for the ZFC
sample, in a field of 6T (see figure 1). The main difference between the FC
and the ZFC samples is that for the former there is a finite but
small DOS at $E_F$ even down to the lowest $T$ which stabilizes
the metallic state as seen from the $\rho$ vs. $T$ curves for the FC sample.
This is a clear indication that the CO state can only be
stabilized in the presence of a gap in the DOS. Even a small DOS at $E_F$
can stabilize the metallic phase as in the FC sample in a field of 6T.
Specific heat measurements have shown 
\begin{figure}
\centerline{
\psfig{figure=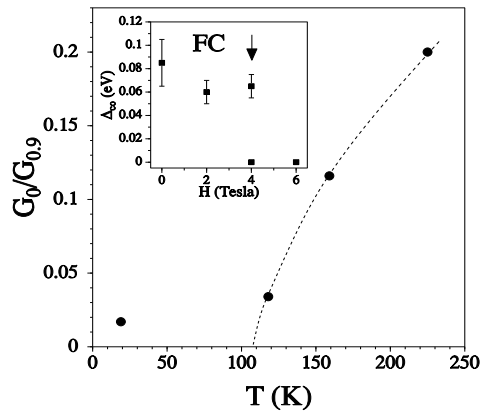,width=7.0cm,height=6.5cm,clip=}
}
\caption{
The variation of $G_0/G_{0.9}$ as the temperature 
is reduced in an applied
field of 6 T. The dotted line is a guide to the eye (details in text).
The inset shows the
variation of $\Delta_{CO}$ with magnetic field
at 25 K of the sample field cooled in 6 T.
Note the two-phase behavior at 4 T as marked by the
arrow.}
\end{figure} 
\vspace{0.2cm}
that the DOS at $E_F$ is very
small for a field cooled sample of the CO compound 
La$_{0.5}$Ca$_{0.5}$MnO$_3$ which has a metallic 
behavior of the $\rho$ vs. $T$ ~\cite{vera}. When 
the field is reduced to  a lower value (staying at low temperatures) the 
charge ordering sets in at a field of 4T as per the resistivity data (see  
figure 2). But
the resistivity at the lowest temperature is less than that of 
of the ZFC sample. This may imply that in 
the FC sample the CO is incomplete even when the field is reduced to 
zero.
The 
variation 
of 
$\Delta_{CO}$ as the field is reduced in a FC sample at low temperatures 
is shown in the inset of
figure 5. One can clearly see the onset of the CO gap at 
$H\leq$ 4T. Interestingly the value of $\Delta_{CO}$ at $H$=0 is much less 
than that seen in the ZFC sample. 
This also points to an incomplete CO in the 
FC sample, when the field is removed.

In the FC sample we make yet another observation. When the field is reduced 
at low temperature and the CO state sets in, there is a clear existence of 
two phases as we can make out from the TS data as shown in 
the inset of figure 5. At a field of 4 T the measured gap is zero 
at some positions on the sample and finite at other positions. This behavior
is denoted by the two values of $\Delta_{CO}$ at $H$=4T and
is marked by an arrow.

Our experiment in both FC and ZFC samples clearly establishes that the 
stability of the CO state under a magnetic field depends on whether there 
is a gap in the DOS at 
$E_F$. In a recent experiment on the compound                            
(NdLa)$){0.5}$Ca$_{0.5}$MnO$_{3}$ we found that a lattice instability can  
lead to a collapse of the incipient CO state
~\cite{ndlaca}. In that case also we find that a 
gap in the DOS at $E_F$ closes when the CO state collapses. We can thus 
conclude that in these compounds a gap in the DOS at $E_F$ 
($\Delta_{CO}$) is a necessary condition for the stability of the CO   
state.

When a CO state melts in a magnetic field, 
$\rho$ changes substantially. This can arise either from 
changes 
in 
the mobility, $\mu$ or in the the free carrier density,$n$ or both. In the 
specific 
case of Nd$_{0.5}$Sr$_{0.5}$MnO$_{3}$ the CO state is also a spin ordered 
AFM state. In a magnetic field it undergoes a metamagnetic transition and 
becomes ferromagnetic. In the ferromagnetic state (which is stabilized by 
double exchange interaction) the spin alignment will definitely lead to 
an 
enhancement of $\mu$ leading to a decrease in $\rho$. However, if there is 
a gap in the DOS at $E_F$ which closes in the magnetic field, it will 
lead 
to an enhancement of $n$ when the CO state melts. This in turn will 
reduce 
the $\rho$ substantially. Our experiment suggests that 
a substantial contribution in 
the 
reduction in $\rho$ on melting of the CO state arises from the gap 
closing. Independent experimental tests like Hall measurement can establish 
whether there is indeed an enhancement of $n$ when the CO state melts in a 
magnetic field.
 
To summarize we have shown that when a CO state melts in a magnetic field 
the gap in the DOS at $E_F$ collapses. We have concluded that a gap in the 
DOS at $E_F$ is needed for stability of the CO state. As result of the 
gap closing the free carrier density $n$ should increase on melting
of the CO state, leading 
to a reduction in $\rho$ by orders of magnitude.
The magnitude of $\Delta_{CO}$ 
is large compared to the thermal scale $k_{B}T_{CO}$ . At
$T/T_{CO}\approx$0.7 the CO state melts in a field of $\approx$ 4 T.
The scale of magnetic energy required to
collapse the CO gap is an order of magnitude less
than the scale of $\Delta_{CO}$. In our opinion this anomaly of the energy
scales remains an open and relevant 
issue and it is a feature characteristic of the
charge ordering phenomenon in manganites.

One of us (AKR) wants to thank DAE (BRNS), Government of India, for a 
sponsored project.

%\end{multicols}

\end{document}